\documentclass[a4paper,showpacs,prb,twocolumn]{revtex4}
\usepackage{amsfonts}
\usepackage{graphicx}
\usepackage{color}
\usepackage{amsmath}
\usepackage{amssymb}
\usepackage{latexsym}
\usepackage{psfrag}

\begin{document}

\title{Magnetosubband and edge state structure in cleaved-edge overgrown
quantum wires}
\author{S. Ihnatsenka and I. V. Zozoulenko}
\affiliation{Solid State Electronics, Department of Science and Technology (ITN), Link%
\"{o}ping University, 60174 Norrk\"{o}ping, Sweden}
\date{\today }

\begin{abstract}
We provide a systematic quantitative description of the structure
of edge states and magnetosubband evolution in hard wall quantum
wires in the integer quantum Hall regime. Our calculations are
based on the self-consistent Green's function technique where the
electron- and spin interactions are included within the density
functional theory in the local spin density approximation. We
analyze the evolution of the magnetosubband structure as magnetic
field varies and show that it exhibits different features as
compared to the case of a smooth confinement. In particularly, in
the hard-wall wire a deep and narrow triangular potential well (of
the width of magnetic length $l_B$) is formed in the vicinity of
the wire boundary. The wave functions are strongly localized in
this well which leads to the increase of the electron density near
the edges. Because of the presence of this well, the subbands
start to depopulate from the central region of the wire and remain
pinned in the well region until they are eventually pushed up by
increasing magnetic field. We also demonstrate that the spin
polarization of electron density as a function of magnetic field
shows a pronounced double-loop pattern that can be related to the
successive depopulation of the magnetosubbands. In contrast to the
case of a smooth confinement, in hard-wall wires the compressible
strips do not form in the vicinity of wire boundaries and spatial
spin separation between spin-up and spin-down states near edges is
absent.
\end{abstract}

\maketitle

\section{Introduction}

Recent advances in fabrication of low-dimensional structures allow one to
create quantum wires with a hard-wall potential confinement. The available
technologies include implantation-enhanced interdiffusion technique \cite%
{Cibert} developed more than 20 years ago. Using this technique Prins
\textit{et al.} \cite{Prins} demonstrated a potential jump at a
hetero-interface GaAs-AlGaAs over only 8 nm distance. The molecular beam
epitaxy double-growth technique \cite{Pfeiffer1} (often referred to as a
cleaved-edge overgrowth) since early 1990-th has become one of the most
widely-used techniques for fabrication of quantum wires \cite%
{Yacoby,Yacoby2,Motoshita} and two-dimensional electron gases (2DEGs) \cite%
{Huber} with an essentially hard wall confinement with the atomic precision.
Quantum wires with a steep confinement can also be fabricated by overgrowth
on patterned GaAs(001) substrates using molecular beam epitaxy \cite{Shitara}%
.

For theoretical description of the quantum Hall effect in quantum wires, a
concept of edge states is widely used\cite{Halperin}. In a naive
one-electron picture a position of the edge states are determined by the
intersection of the Landau levels (bent by the bare potential) with the
Fermi energy, and their width is given by a spatial extension of the wave
function, which is of the order of the magnetic length $l_{B}=\sqrt{\frac{%
\hbar }{eB}}$. For a smooth electrostatic confinement that varies
monotonically throughout the cross-section of a wire, Chklovskii \textit{at
al.}\cite{Chklovskii} have shown that electrostatic screening in strongly
modifies the structure of the edge states giving rise to interchanging
compressible and incompressible strips. The electrons populating the
compressible strips screen the electric field, which leads to a metallic
behavior when the electron density is redistributed (compressed) to keep the
potential constant. The neighboring compressible strips are separated from
each other by insulator-like incompressible strips corresponding to the
fully filed Landau levels with a constant electron density.

A number of studies of quantum wires with a smooth confinement have been
reported during the recent decade\cite%
{Kinaret,Ando_1993,Dempsey,Tejedor,Gerhardts_1994,Tokura,Stoof,%
Ferconi_1995,Takis,Ando_1998,Schmerek,Gerhardts_2004}
addressing the problem of electron-electron interaction beyond Chklovskii
\textit{at al.}'s\cite{Chklovskii} electrostatic treatment. A particular
attention has been paid to spin polarization effects in the edge states \cite%
{Kinaret,Dempsey,Tokura,Takis,Stoof,Ihnatsenka,Ihnatsenka2}. It has been
demonstrated that the exchange and correlation interactions dramatically
affect the edge state structure in quantum wires bringing about
qualitatively new features in comparison to a widely used model of spinless
electrons. These include spatial spin polarization of the edge states\cite%
{Dempsey,Ihnatsenka2}, pronounced $1/B$-periodic spin polarization of the
electron density\cite{Ihnatsenka}, modification and even suppression of the
compressible strips\cite{Ihnatsenka2} and others. It should be stressed that
all the above-mentioned studies addressed the case of a soft confinement
corresponding to e.g. a gate-induced depletion when the Borh radius is much
smaller that the depletion length. In fact, Huber \textit{et al.} have
recently presented experimental evidence that widely used concept of
compressible/incompressible strips\cite{Chklovskii} does not apply to the
case of a sharp-edge 2DEG. At the same time the rigorous theory for
edge-state structure in hard-wall quantum wires accounting for
electron-electron interaction and spin effects has not been reported yet.
Such a theory is obviously required for a detailed analysis of recent
experiments on cleaved-edge overgrown sharp-edge wires and 2DEGs \cite%
{Prins,Pfeiffer1,Huber,Yacoby,Yacoby2,Motoshita,Shitara}.

Motivated by the above-mentioned experimental studies, in this paper we
present a detailed theory of magnetosubband and edge state structure in
quantum wires with a hard wall confinement taking into account
electron-electron interaction including exchange and interaction effects. We
employ an efficient numerical tool based on the Green's function technique
for self-consistent solution of the Schr\"{o}dinger equation in the
framework of the density functional theory (DFT) in the local spin density
approximation (LSDA)\cite{ParrYang}. The choice of DFT+LSDA for description
of many-electron effects is motivated, on one hand, by its efficiency in
practical implementation within a standard Kohn-Sham formalism \cite{Kohn},
and, on the other hand, by an excellent agreement between the DFT+LSDA and
the exact diagonalization \cite{Ferconi} and the variational Monte-Carlo
calculations\cite{Rasanen,QDOverview} performed for few-electron quantum
dots. We will demonstrate below that edge state structure of the hard wall
quantum wire is qualitatively different from that of the soft-wall wire. We
will discuss how the spin-resolved subband structure, the current densities,
the confining potentials, as well as the spin polarization in the hard wall
quantum wire evolve when an applied magnetic field varies.

The paper is organized as follow. In Sec. II we present a formulation of the
problem, where we define the geometry of the system at hand and outline the
self-consistent Kohn-Sham scheme within the DFT+LSDA approximation. In Sec.
III we present our results for a hard wall quantum wire calculated within
Hartree and DFT+LSDA approximations, where we distinguish cases of wide and
narrow wires. Section IV contains our conclusions.

\section{Model}

We consider a quantum wire which is infinitely long in the $x$-direction and
is confined by a hard-wall potential in the $y$-direction, see Fig. \ref%
{fig:vconf}. 
\begin{figure}[!tb]
\includegraphics[scale=0.8]{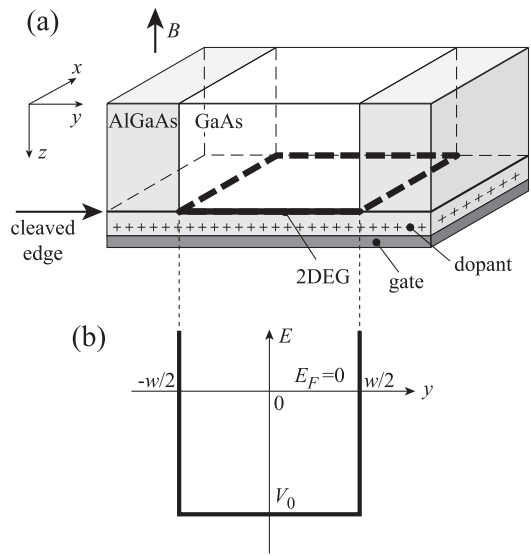} 
\caption{(Color online). (a) A schematic illustration of a cleaved-edge
overgrown quantum wire and (b) a corresponding hard-wall confinement
potential.}
\label{fig:vconf}
\end{figure}

The magnetic field is applied perpendicular to the $xy$-plane. We set the
Fermi energy $E_{F}=0$. A bottom of the confining potential is flat and
situated at $E=V_{0}$. We limit ourself to a typical case when only one
subband is occupied in the transverse $z$-direction\cite{Huber} such that
electron motion is confined to the $xy$-plane. The Hamiltonian of the wire
reads $H=\sum_{\sigma }H^{\sigma }$,
\begin{equation}
H^{\sigma }=H_{0}+V_{0}+V_{eff}^{\sigma }(y)+g\mu _{b}B\sigma ,
\label{Hamiltonian}
\end{equation}%
where $H_{0}$ is the kinetic energy in the Landau gauge,
\begin{equation}
H_{0}=-\frac{\hbar ^{2}}{2m^{\ast }}\left\{ \left( \frac{\partial }{\partial
x}-\frac{eiBy}{\hbar }\right) ^{2}+\frac{\partial ^{2}}{\partial y^{2}}%
\right\} ,  \label{2}
\end{equation}%
where $\sigma =\pm \frac{1}{2}$ describes spin-up and spin-down states, $%
\uparrow $, $\downarrow $, and $m^{\ast }=0.067m_{e}$ is the GaAs effective
mass. The last term in Eq. (1) accounts for Zeeman energy where $\mu _{b}=%
\frac{e\hbar }{2m_{e}}$ is the Bohr magneton, and the bulk $g$ factor of
GaAs is $g=-0.44$. The effective potential, $V_{eff}(y)$ within the
framework of the Kohn-Sham density functional theory reads \cite{Kohn},
\begin{equation}
V_{eff}^{\sigma }(y)=V_{H}(y)+V_{ex}^{\sigma }(y),  \label{Veff}
\end{equation}%
where $V_{H}(y)$ is the Hartree potential due to the electron density $%
n(y)=\sum_{\sigma }n^{\sigma }(y)$ (including the mirror charges) \cite%
{Ihnatsenka},
\begin{equation}
V_{H}(y)=-\frac{e^{2}}{4\pi \varepsilon _{0}\varepsilon _{r}}\int dy^{\prime
}n(y^{\prime })\ln \frac{\left( y-y^{\prime }\right) ^{2}}{\left(
y-y^{\prime }\right) ^{2}+4b^{2}}.  \label{Hartree}
\end{equation}%
with $2b$ being the distance from the electron gas to the mirror charges (we
choose $b$=60 nm). For the exchange and correlation potential $V_{xc}(y)$ we
utilize the widely used parameterization of Tanatar and Cerperly \cite{TC}
(see Ref. \onlinecite{Ihnatsenka} for explicit expressions for $V_{xc}(y)$).
This parameterization is valid for magnetic fields corresponding to the
filling factor $\nu >1$, which sets the limit of applicability of our
results. The spin-resolved electron density reads
\begin{equation}
n^{\sigma }(y)=-\frac{1}{\pi }\Im \int dE\,G^{\sigma }(y,y,E)f_{FD}(E-E_{F}),
\label{density}
\end{equation}%
where $G^{\sigma }(y,y,E)$ is the retarded Green's function corresponding to
the Hamiltonian (\ref{Hamiltonian}) and $f_{FD}(E-E_{F})$ is the Fermi-Dirac
distribution function. The Green's function of the wire, the electron and
current densities are calculated self-consistently using the technique
described in detail in Ref. \onlinecite{Ihnatsenka}.

The current density for a mode $\alpha $ is calculated as \cite{Ihnatsenka}
\begin{equation}
J_{\alpha }^{\sigma }(y)=\frac{e^{2}}{h}V\int dE\frac{j_{\alpha }^{\sigma
}(y,E)}{v_{\alpha }^{\sigma }}\left( -\frac{\partial f\left( E-E_{F}\right)
}{\partial E}\right) ,  \label{J}
\end{equation}%
with $v_{\alpha }^{\sigma }$ and $j_{\alpha }^{\sigma }(y,E)$ being
respectively the group velocity and the quantum-mechanical particle current
density for the state $\alpha $ at the energy $E$, and $V$ being the applied
voltage.

We also calculate a thermodynamical density of states ($TDOS$) defined
according to \cite{Davies,Manolescu}
\begin{equation}
TDOS^{\sigma }=\int dE\,\rho ^{\sigma }(E)\left( -\frac{\partial
f_{FD}(E-E_{F})}{\partial E}\right) ,  \label{TDOS}
\end{equation}%
where the spin-resolved density of states $\rho ^{\sigma }(E)$ is given by
the Green function\cite{Datta},
\begin{equation}
\rho ^{\sigma }(E)=-\frac{1}{\pi }\Im \int dy\,\,G^{\sigma }(y,y,E).
\label{dos}
\end{equation}%
The $TDOS$ reflects a structure of the magnetosubbands near the Fermi energy
and it can be accessible via magneto-capacitance \cite{Weiss} or
magnetoresistance \cite{Berggren} measurements. Indeed, a compressible strip
corresponds to a flat (dispersionless) subband pinned at $E_{F}$. In this
case $\rho ^{\sigma }(E)$ is high at $E\approx E_{F}$ and such the subband
strongly contributes to $TDOS$. In contrast, in an incompressible strip,
subbands are far away from $E_{F}$ and do not contribute to $TDOS$. Thus $%
TDOS$ is proportional to the area of the compressible strips. This area is
maximal when the strip if formed in the middle of a quantum wire. In this
case the backscattering between opposite propagating states is maximal,
which corresponds to peaks in the longitudinal resistance $R_{xx}$ (seen as
the Shubnikov-De Haas oscillations) \cite{Berggren, Hwang, Beenakker}. In
magneto-capacitance experiments\cite{Weiss,Beenakker} the compressible
strips are viewed as capacitor plates and therefore the measured
magnetocapacitance is related to the width of these strips. Thus the peaks
in the $TDOS$ are manifest themselves in both $R_{xx}$ and capacitance peaks.

\section{Results and discussion}

In what follows we shall distinguish between cases of a wide quantum wire
whose half-width $\frac{w}{2}$ exceeds the magnetic length $l_{B},$ and a
narrow wire with a width $\frac{w}{2}\lesssim $ $l_{B}.$
\begin{figure*}[!tb]
\includegraphics[scale=1.1]{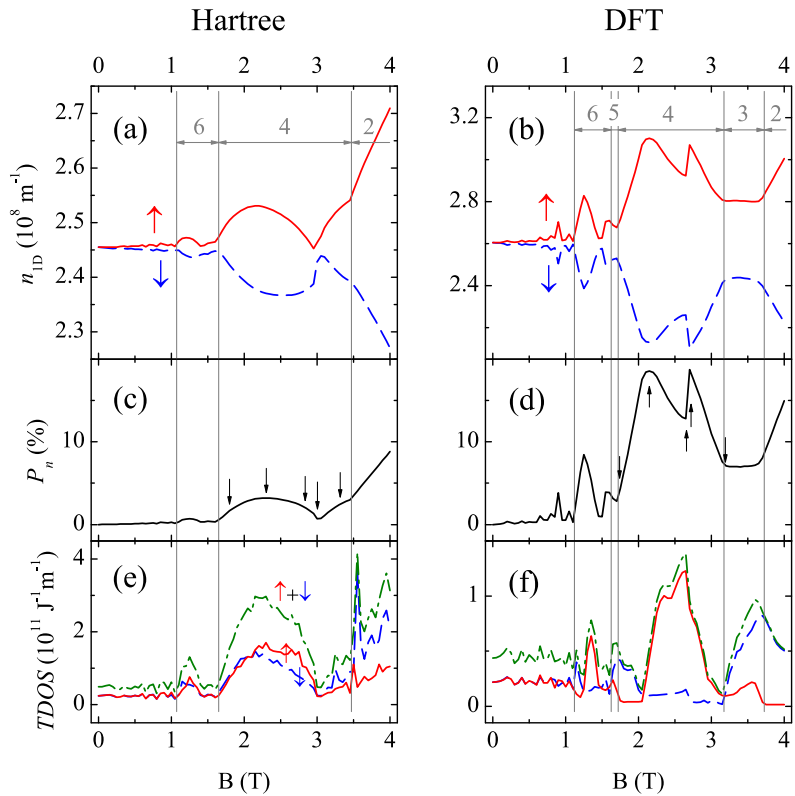} 
\caption{(Color online). (a),(b) One-dimensional charge density
for the spin-up and spin-down electrons, $n_{1D}^{\uparrow},\
n_{1D}^{\downarrow}$; (c),(d) the spin polarization of the charge
density, $P_{n}=\frac{n_{1D}
^{\uparrow}-n_{1D}^{\downarrow}}{n_{1D}^{\uparrow}+n_{1D}^{\downarrow}}$,
(g),(h) the $TDOS$ for spin-up and spin-down electrons and the
total $TDOS$ within the Hartree approximation and the DFT
approximation (first and second columns, respectively). The number
of subband is indicated in (a),(b). Arrows in (c) and (d) indicate
the magnetic field corresponding to the magnetosubband structure
shown in Figs. \protect\ref{fig-wide2d-hartree} and
\protect\ref{fig-wide2d-dft}. The width of the wire is $w=300$ nm
and the depth is $V_{0}=-0.1$ eV. Temperature $T=1$ K.}
\label{fig-wide1d}
\end{figure*}

\subsection{Wide hard wall quantum wire $\frac{w}{2}>l_{B}$}

Let us consider a hard wall quantum wire of the width $w=300$ and $V_{0}=-0.1
$ eV. With these parameters the wire has $N\sim 20$ spin-resolved occupied
subbands at zero magnetic field, and the sheet electron density in its
center is $n_{2D}\approx 1.5\cdot 10^{15}$ m$^{-2}$ (as calculated
self-consistently in both Hartree and DFT approximations).

\paragraph{Hartree approximation}

We start our analysis of the edge state- and magnetosubband structure from
the case of the Hartree approximation (when the exchange and correlation
interactions are not included in the effective potential). The Hartree
approximation gives the structure of the compressible/incompressible strips
which serves as a basis for understanding of the effect of the exchange and
correlation within the DFT approximation\cite{Ihnatsenka,Ihnatsenka2}.
\begin{figure*}[!ht]
\includegraphics[scale=0.95]{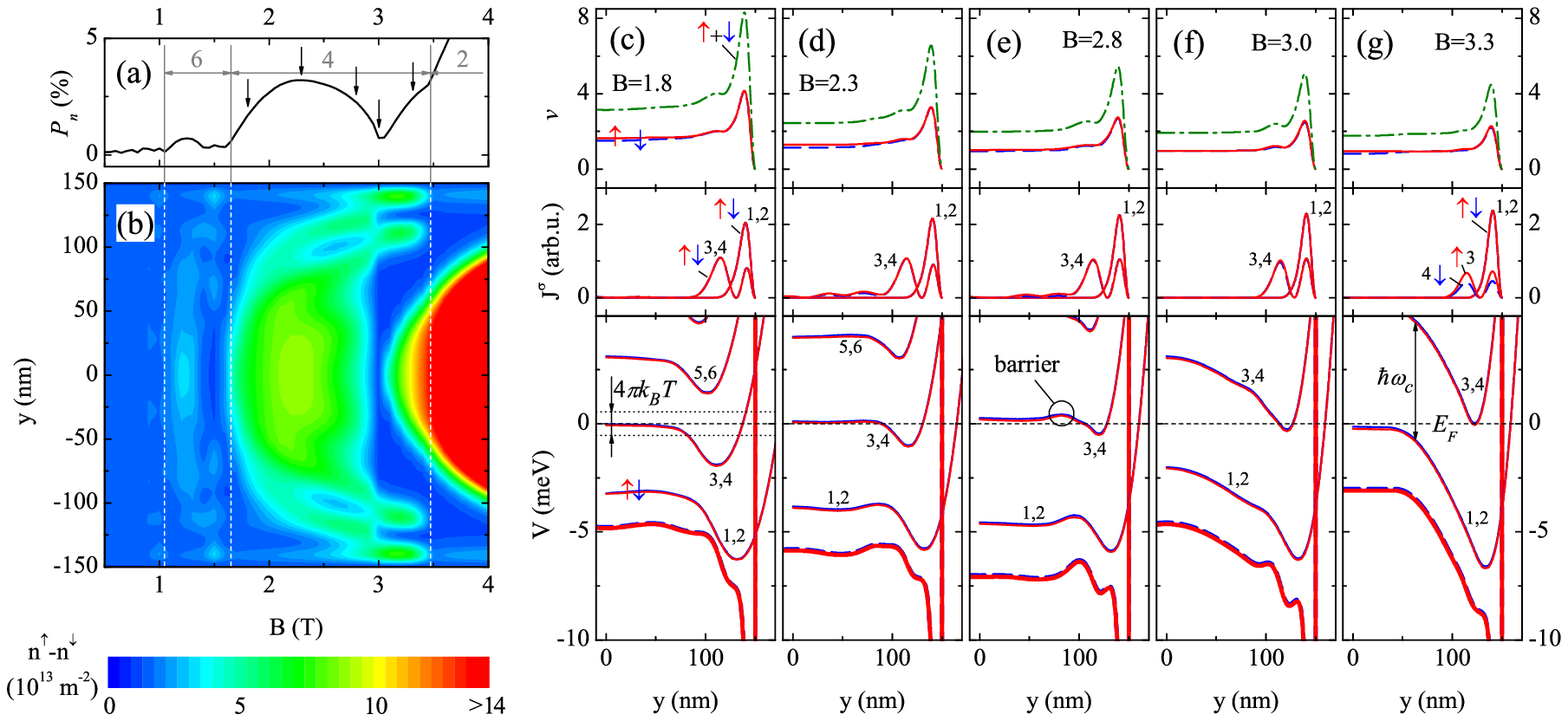} 
\caption{(Color online). (a) Spin polarization of the charge
density as a function of $B $ calculated within the Hartree
approximation (the same as Fig. \protect\ref{fig-wide1d}(c)). (b)
Spatially resolved difference in the electron density
$n^{\uparrow}(y)-n^{\downarrow}(y)$. (c)-(g) The subband structure
for magnetic fields indicated by arrows in (a). Upper panel:
electron density profiles (local filling factors)
$\protect\nu(y)=n(y)/n_{B}$ for spin-up and spin-down electrons;
middle panel: the current density distribution for spin-up and
spin-down electrons; lower panel: magnetosubband structure for
spin-up and spin-down electrons. Fat solid and dashed lines
indicate the total confining potential for respectively spin-up
and spin-down electrons. The width of the wire is $a=300 $ nm and
depth is $V_{0}=-0.1$ eV. Temperature $T$=1 K.}
\label{fig-wide2d-hartree}
\end{figure*}

Figure \ref{fig-wide1d}(a) shows the 1D electron density $n_{1D}^{\sigma
}=\int n^{\sigma }(y)dy$ for the spin-up and spin-down electrons in the
quantum wire. The pronounced feature of this dependence is a characteristic
loop pattern of the charge density polarization, $P_{n}=\frac{%
n_{1D}^{\uparrow }-n_{1D}^{\downarrow }}{n_{1D}^{\uparrow
}+n_{1D}^{\downarrow }}$, see Fig. \ref{fig-wide1d}(b) . Figure \ref%
{fig-wide1d} also indicates a number of magnetosubbands $N$ populated at a
given $B$. The number of subbands is always even such that spin-up and
spin-down subbands depopulate practically simultaneously. This is because
the spin polarization within the Hartree approximation is driven by Zeeman
splitting only, which is small in the field interval under consideration. A
comparison of Figs. \ref{fig-wide1d}(a),(c),(e) demonstrates that the spin
polarization as well as the  $TDOS$ are directly related to the
magnetosubband structure. Note that a similar loop-like behavior of the spin
polarization is also characteristic for a split-gate wire with a smooth
confinement\cite{Ihnatsenka}. For the latter case the polarization
calculated in the Hartree approximation drops practically to zero when the
subbands depopulate (see Fig. 4 in Ref. \onlinecite{Ihnatsenka}). In
contrast, in the case of the hard wall confinement, the polarization loops
exhibit more complicated pattern: the polarization does not drop to zero
when the subbands depopulate, and, in addition, the polarization curves show
a double loop-like pattern with an additional minimum (e.g. at $B\approx 1.5$%
T, 3T in Fig. \ref{fig-wide1d} (a),(c)). In order to understand the origin
of this behaviour let us analyze the evolution of the subband structure as
the applied magnetic field varies. Let us concentrate at the field interval
1.65 T $\lesssim B\lesssim $ 3.5 T when the subband number $N=4$.

Figure \ref{fig-wide2d-hartree}(b) shows the spatially resolved difference
in the electron density $n^{\uparrow }(y)-n^{\downarrow }(y)$ as a function
of $B$. The electron density is mostly polarized in the inner region of the
quantum wire. For certain ranges of magnetic fields the electron density
shows a strong polarization in the boundary regions, which are separated
from the polarized inner region by wide unpolarized strips (e.g. for $3\text{%
T}\lesssim B\lesssim 3.5\text{T}$). We will show below that this feature
reflects the peculiarities of the magnetosubband structure for the case of
the hard wall confinement. Figure \ref{fig-wide2d-hartree}(c) shows the
electron density profiles (local filling factors) $\nu (y)=n(y)/n_{B}$ $%
(n_{B}=eB/h)$, the current densities $J^{\sigma }(y)$ and the magnetosubband
structure for the magnetic field $B=1.8$ T. At this field a wide
compressible strip due to electrons belonging to the subbands $N=3,4$ is
formed in the middle of the wire. (Following Suzuki and Ando\cite{Ando_1998}
we define the width of the compressible strips within the energy window $%
|E-E_{F}|<2\pi kT$ corresponding to the partial occupation of the subbands
when $f_{FD}<1$; this energy window is indicated in Fig. \ref%
{fig-wide2d-hartree} (c)). Partial subband occupation combined with Zeeman
splitting of energy levels results in different population for spin-up and
spin-down electrons (i.e. in the spin polarization of the electron density).

Close to the wire edges the total potential exhibits a narrow and deep
triangular well. The formation of the triangular well is also reflected in
the structure of the magnetosubbands that show triangular wells near the
wire edges. Presence of these triangular wells is a distinctive feature of
the hard-wall confinement (it is absent for the case of a smooth confinement
in the split-gate wires\cite{Ando_1993,Ando_1998,Ihnatsenka,Ihnatsenka2}).
The wave functions for all subbands are strongly localized in these wells,
with the extension of the wave functions being of the order of the magnetic
length $l_B$. Because of steepness of the potential walls, the wave
functions are not able to screen the confining potential, and compressible
strips can not form near the wire boundary. This is in a stark contrast to
the case of a split-gate wire where the compressible strips near edges are
formed for a sufficiently smooth confinement\cite%
{Chklovskii,Ando_1998,Ihnatsenka,Ihnatsenka2}. The electron density near the
wire boundaries does not show any spin polarization. This is because the
bottom of the potential well lies far below the Fermi energy. As a result,
both spin-up and spin-down states localized in the quantum well are
completely filled ($f_{FD}=1$) and the spin polarization is absent.

When a magnetic field increases the compressible strip in the middle of the
wire widens. This is accompanied by increase of both the spin polarization
and the $TDOS$ as shown in Figs. \ref{fig-wide1d}(c),(e). At $B=2.3$T the
polarization reaches maximum $P_{n}=3$ \% which corresponds to the maximum
width of the compressible strip in the central part of the wire, see Fig. %
\ref{fig-wide2d-hartree}(d). With further increase of the magnetic field 3rd
and 4th subbands in the central part of the wire are pushed up, see Fig. \ref%
{fig-wide2d-hartree}(e). Their population decreases according to the
Fermi-Dirac distribution and, consequently, the spin polarization
diminishes. At the same time, fully occupied parts of 3rd and 4th subbands
(forming a triangular well near the wire boundaries) are pushed up and got
pinned at the Fermi energy. This is accompanied by a formation of a
potential barrier at the distance of the wave function extent $\sim l_{B}$
from the wire edges, see Fig. \ref{fig-wide2d-hartree}(e). The whole area
occupied by subbands 3 and 4 becomes divided by non-populated region within
the barrier where the subbands lie above $E_{F}$ (i. e. $f_{FD}=0$).

When a magnetic field slightly increases from $B=2.8$T to $B=3.0$T the
magnetosubband structure undergoes significant changes. A middle part of the
3rd and 4th subbands is abruptly pushed up in energy. The incompressible
strip emerges here due to 1st and 2nd fully occupied subband lying well
below $E_{F}$, Fig. \ref{fig-wide2d-hartree}(f). As a result the spin
polarization decreases and the fist polarization loop closes down at $%
B\approx 3$T, see \ref{fig-wide2d-hartree}(a). Note that $P_{n}$ does not
drop to zero because of a finite polarization at the boundaries where the
3rd and 4th subband bottoms are still pinned at the Fermi energy, see Fig. %
\ref{fig-wide2d-hartree}(b),(f). As magnetic field increases the second
polarization loop starts to form at $B\approx 3$T due to 1st and 2nd
subbands that get pinned to $E_{F}$ in the middle of the wire (Fig. \ref%
{fig-wide2d-hartree}(g)). In addition, 3rd and 4th subbands that are pinned
to $E_{F}$ near the wire boundaries also contribute to spin polarization.
These subbands become completely depopulated at $B=3.5$ T. Further increase
of the magnetic field causes the compressible strip in the middle to widen.
The spin polarization $P_{n}$ grows linearly until the second subband
becomes depopulated.

Note that the above scenario of the subband depopulation in quantum wires
with a hard wall confinement is qualitatively different from that one of the
smooth confinement. In the former case, because of the presence of the deep
triangular well near the wire boundaries, the subbands start to depopulate
from the central region of the wire and remain pinned in the well region
until they are eventually pushed up by magnetic field. In contrast, in the
case of a smooth confinement, the subband always depopulate from the edges,
such that a compressible strip in the middle of the wire gradually decreases
until it completely disappears when the whole subband is pushed up above the
Fermi energy\cite{Ihnatsenka,Ihnatsenka2}.

The spatial current distribution stays practically the same throughout the
magnetosubband evolution, see the central panel in Figs. \ref%
{fig-wide2d-hartree}(c)-(g). This is due to a strong localization of
electrons in the triangular potential well. The spatial spin separation
between spin-up and spin-down states is always equal to zero, which is also
the case for a split-gate wire in the Hartree approximation \cite%
{Ihnatsenka,Ihnatsenka2}.

Finally, within the Hartree approximation the $TDOS$ shows a behavior
similar to the spin polarization of the electron density $P_{n}$, compare
Fig. \ref{fig-wide1d}(e) and \ref{fig-wide1d}(c). This is because the spin
polarization is primarily caused by electrons in the compressible strips,
and the $TDOS,$ as discussed in the previous section, is proportional to the
width of these strips.

\begin{figure*}[!ht]
\includegraphics[scale=0.95]{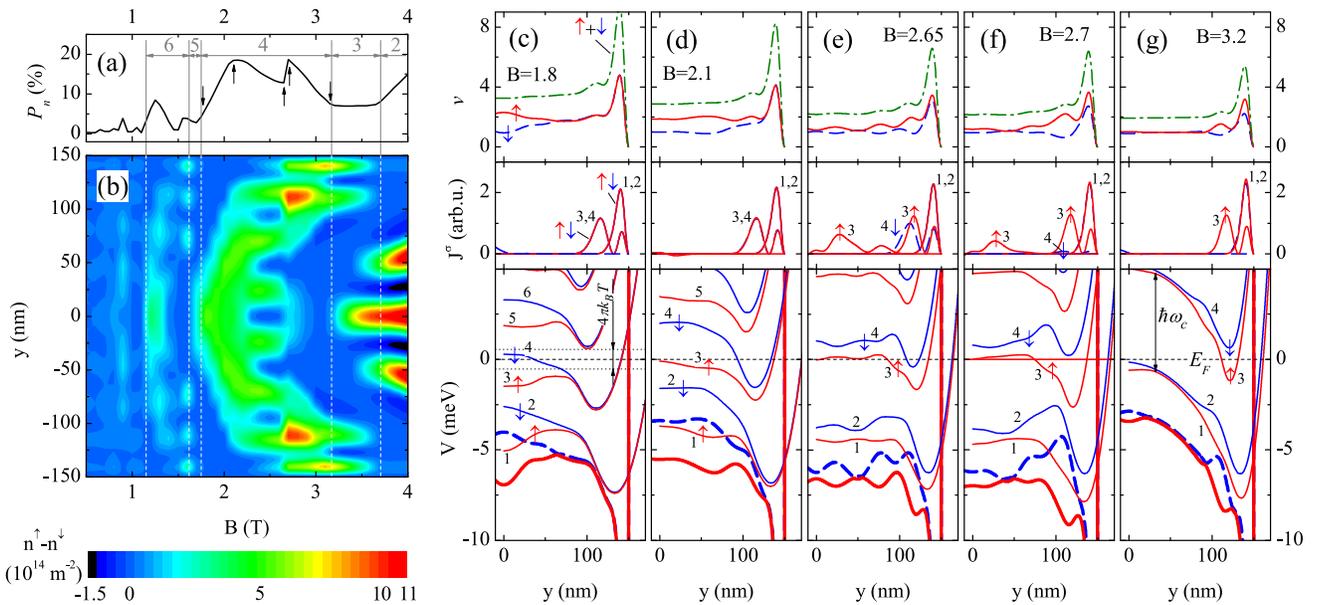}
\caption{(Color online). (a) Spin polarization of the charge
density as
a function of $B$ calculated within the DFT approximation (similar to Fig.
\protect\ref%
{fig-wide1d}(d)). (b) Spatially resolved difference in the electron density $%
n^{\uparrow}(y)-n^{\downarrow}(y)$. (c)-(g) The subband structure
for magnetic fields indicated by arrows in (a). Upper panel:
electron density profiles (local filling factors)
$\protect\nu(y)=n(y)/n_{B}$ for spin-up and spin-down electrons;
middle panel: the current density distribution for spin-up and
spin-down electrons; lower panel: magnetosubband structure for
spin-up and spin-down electrons. Fat solid and dashed lines
indicate the total confining potential for respectively spin-up
and spin-down electrons. The width of the wire is $a=300$ nm and
depth is $V_{0}=-0.1$ eV. Temperature $T=1$K.}
\label{fig-wide2d-dft}
\end{figure*}

\paragraph{DFT approximation}

The exchange and correlation interactions bring qualitatively new features
to the magnetosubband structure in comparison to the Hartree approximation.
Figures \ref{fig-wide1d}(b),(d),(f) show the 1D electron density, the number
of subbands, the spin polarization and the $TDOS$ calculated within DFT
approximation. There are several major differences in comparison to the
Hartree case. First, the spin polarization of the electron density also
shows a pronounced loop pattern. However, for a given magnetic field the
spin polarization in the quantum wire calculated on the basis of the DFT
approximation is much higher in comparison to the Hartree approximation (by
a factor 5-10). Second, the exchange interaction lifts subband degeneration,
such that the subbands depopulate one by one. Third, the $TDOS$ reveals
peaks which are attributed to different spin species.

Before we proceed to analysis of the magnetosubband structure within the DFT
approximation, it is instrumental to outline the effect of the exchange
interaction on the subband spin splitting. Within the Hartree approximation
the subbands are practicably degenerate because the Zeeman splitting is very
small in the magnetic field interval under investigation. In contrast, the
exchange interaction included within the DFT approximation causes the
separation of the subbands which magnitude can be comparable to the Landau
level spacing $\hbar \omega $. Indeed, the exchange potential for spin-up
electrons depends on the density of spin-down electrons and vice versa\cite%
{ParrYang,TC,Ihnatsenka}. In the compressible region the subbands are only
partially filled (because $f_{FD}<1$ in the the window $|E-E_{F}|\lesssim
2\pi kT$), and, therefore, the population of the spin-up and spin-down
subbands can be different. In the DFT calculation, this population
difference (triggered by Zeeman splitting) is strongly enhanced by the
exchange interaction leading to different effective potentials for spin-up
and spin-down electrons and eventually to the subband spin splitting. Below
the Fermi energy $E\lesssim E_{F}-2\pi kT$ the subbands remain degenerate
because they are fully occupied ($f_{FD}=1$). As a result, the corresponding
spin-up and spin-down densities are the same, hence the exchange and
correlation potentials for the spin-up and spin-down electrons are equal, $%
V_{xc}^{\uparrow }(y)=V_{xc}^{\downarrow }(y)$.

In order to understand the effect of the exchange-correlation interactions
on evolution of the magnetosubband structure, let us concentrate on the same
field interval as discussed in the case of the Hartree approximation, 1.8 T $%
\lesssim B\lesssim $ 3.7 T. A comparison between Fig. \ref{fig-wide2d-dft}
and Fig. \ref{fig-wide2d-hartree} demonstrates that evolution of the
magnetosubband structure calculated within the DFT approximation follows the
same general pattern as for the case of the Hartree approximation. In
particularly, a deep triangular well near the wire boundary develops in the
total confining potential for both spin-up and spin-down electrons. The wave
functions are strongly localized in this well. As a result, similarly to the
Hartree case, the depopulation of the subbands starts from the central
region of the wire. The subbands remain pinned in the well region until they
are eventually pushed up by magnetic field. The major difference from the
Hartree case is that Hartree subbands are practically degenerated and
depopulate together, whereas this degeneracy is lifted by the exchange
interaction such that DFT subbands depopulate one by one. Indeed, Figs. \ref%
{fig-wide2d-dft} (c),(d) showing consecutive depopulation of the subbands 4
and 3 in the central region of the wire can be compared with the
corresponding evolution of the Hartree subbands in Figs. \ref%
{fig-wide2d-hartree} (c),(d). When the magnetic field increases further, 3rd
subband bends upward in the vicinity of the triangular well, compare Fig. %
\ref{fig-wide2d-dft} (e) and Fig. \ref{fig-wide2d-hartree} (e). When
magnetic field reaches $B\approx 2.7$T, 4th spin-down subband becomes
completely depopulated and 3rd spin-up subband is occupied mostly in the
region of the triangular well near the wire boundary, see Fig. \ref%
{fig-wide2d-hartree} (f). This leads to a strong spin polarization near the
boundary which is manifest itself in the additional loop of the polarization
(see Fig. \ref{fig-wide1d} (b), 2.7T$\lesssim B\lesssim $3.2T). Note that
this loop is absent in the Hartree calculations because both 3rd and 4th
subbands are occupied in the well region, such that the spin splitting
between them is small (see Fig. \ref{fig-wide2d-hartree} (f)). Finally, 3rd
subband becomes fully depopulated in the central region, and a compressible
strip starts to form there due to 2nd subband that is pushed upwards,
compare Figs. \ref{fig-wide2d-dft} (g) and Fig. \ref{fig-wide2d-hartree} (g).

Note that similarly to the case of the Hartree approximation, the evolution
of the magnetosubband structure within the DFT approximation described above
qualitatively holds for all other polarization loops.

We also stress that in contrast to the case of a smooth confinement\cite%
{Ihnatsenka,Ihnatsenka2}, in hard-wall wires the compressible strips do not
form in the vicinity of wire boundaries and a spatial spin separation
between spin-up and spin-down states near edges is absent.

Oscillations of the $TDOS$ calculated within the DFT approximation shows
that neighboring peaks belong to different spin species (Fig. \ref%
{fig-wide1d}(f)). In contrast, the Hartree approximation shows that each
single peak includes equal contributions from both species (Fig. \ref%
{fig-wide1d}(e)). It is interesting to note that the oscillation of the $TDOS
$ do not exactly correspond to the subband depopulation. Instead, they
reflect formation of the compressible strip in the middle of the wire due to
spin-up and spin-down electrons which is not directly related to the subband
depopulation (which takes place in the region of the triangular well near
the wire edge).

To conclude this section we note that we analyzed the magnetosubband
structure for a representative sharp-edge quantum wire of 300 nm width. It
is important to stress that all the conclusions presented above (i.e. the
scenario of magnetosubband depopulation and the structure of the edge states
near the wire boundary) hold for an arbitrary sharp-edged quantum wire
provided its length is sufficiently larger than the magnetic length $l_B$.
In particulary, our results can be applied to analysis of an epitaxially
overgrown cleaved edge semi-infinite structure similar to that one studied
in Ref. \onlinecite{Huber}.

\begin{figure*}[!htb]
\includegraphics[scale=1.0]{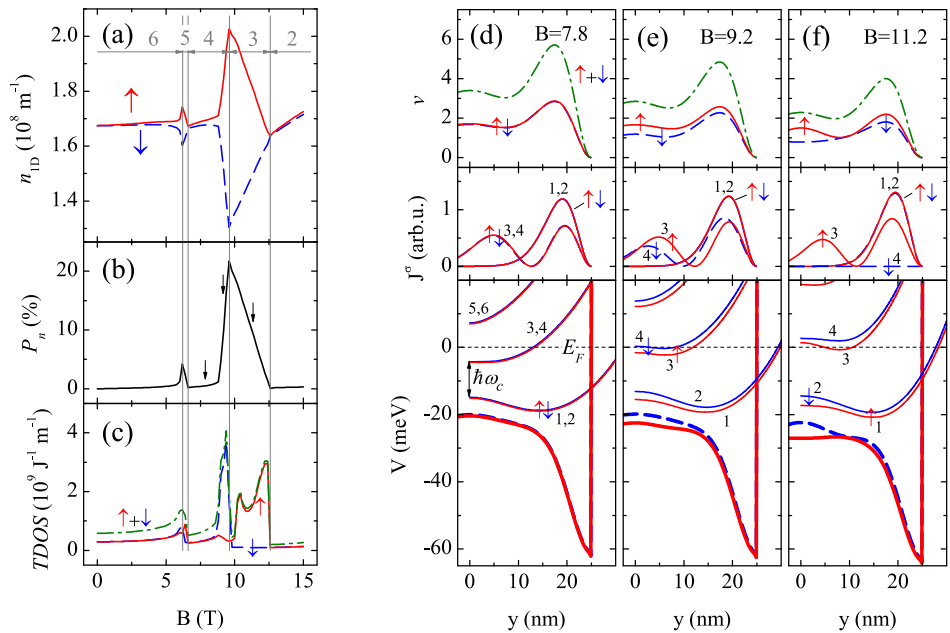}
\caption{(Color online). 1D charge density for spin-up and
spin-down electrons (a), the charge spin polarization (b), the
$TDOS$ for spin-up and spin-down electrons, total $TDOS$ (c) as a
function of $B$ calculated within the DFT approximation for a
narrow wire. (d)-(f) The subband structure for magnetic fields
indicated in (b). Upper panel: electron density profiles (local
filling factors) $\protect\nu(y)=n(y)/n_{B}$; middle panel: the
current density distribution; lower panel: magnetosubband
structure for spin-up and spin-down electrons. Fat solid and
dashed lines indicate the total confining potential for
respectively spin-up and spin-down electrons. The width of the
wire is $a=50$ nm and depth is $V_{0}=-0.2$ eV. Temperature
$T=1$K.} \label{fig-narrow}
\end{figure*}

\subsection{Narrow hard wall quantum wire $\frac{w}{2}\lesssim $ $l_{B}$}

Let us now concentrate on the case of a narrow wire whose half-width is
comparable to the magnetic length. For our analysis we choose the wire of
the width $w$=50 nm and $V_{0}=-0.2$ eV. With these parameters the electron
density at the center of the wire is $n_{2D}\approx6\cdot10^{15}$ m$^{-2}$
and the number of spin-resolved subbands is $N=6$ for $B=0$ T.

Figures \ref{fig-narrow}(a) and (b) show respectively the 1D charge density
and the polarization for spin-up and spin-down electrons calculated within
the DFT approximation. Let us concentrate on the field interval $7\lesssim
B\lesssim 12$, when a number of subbands $3\leq N\leq 4$. In this interval
the spin polarization shows a pronounced single-loop pattern. This is in
contrast to the case of a wide wire that exhibits a double-loop pattern (see
Figs. \ref{fig-wide1d} (a),(b)), where the first loop corresponds to the
subband depopulation in the middle of the wire, whereas the second loops
corresponds to the subband depopulation in the deep triangular well near the
boundary. Note that the width of the this well is of the order of the
extension of the wave function given by the magnetic length $l_{B}$. This
explain a single-loop structure of the polarization curve for the case of a
narrow wire $\frac{w}{2}\lesssim l_{B}$. Indeed, in this case the extension
of the triangular well is comparable to the half-width of the wire, such
that the well extends in the middle region and there is no separate
depopulation for the inner and outer regions of the wire.

The above features of the narrow wire can be clearly traced in the evolution
of the magnetosubbands, see Fig. \ref{fig-narrow}. When 6.5 T $\lesssim
B\lesssim 8.5$ T 3rd and 4th subbands in the middle of the wire are located
beneath $E_{F}-2\pi kT$ and are thus fully occupied. This corresponds to the
formation of the incompressible strip in the middle of the wire such that
the charge densities of spin-up and spin-down electrons are equal (i.e. the
spin polarization is zero). At $B=8.5$T 4th subband reaches $E_{F}-2\pi kT$
and thus becomes partially occupied. As a result, the exchange interactions
generates spin splitting, and the compressible strip due to spin-down
electrons belonging to 4th subband starts to form in the middle of the wire.
Spin polarization grows rapidly until it reaches its maximum $P_{n}=22$ \%.
At this moment 4th subband depopulates and the corresponding compressible
strip disappears. When magnetic field is increased only slightly, 3rd
subband is raised to $E_{F}-2\pi kT$ and the compressible strip due to
spin-up electrons forms in the middle of the wire. Note that formation and
disappearance of the compressible strips due to spin-up and spin-down
electrons is clearly reflected in the $TDOS$, see Fig. \ref{fig-narrow}(c)
which shows peaks belonging to different spin species. With further increase
of $B$ the spin polarization decreases linearly until it vanishes when 3rd
subband fully depopulates.

Magnetosubband evolution calculated within the Hartree approximation (not
shown) qualitatively resembles evolution for the DFT case. In particular,
the spin density polarization follows the same behavior reaching the maximum
value $P_{n}=10$ in the interval $3\leq N\leq 4$. The similarity between the
Hartree and DFT approximations is because of a large Zeeman term for
magnetic field intervals under consideration which causes a relatively
strong Zeeman splitting in the Hartree approximation.

\section{Conclusion}

We provide a systematic quantitative description of the structure of the
edge states and magnetosubband evolution in hard wall quantum wires in the
integer quantum Hall regime. Our calculations are based on the
self-consistent Green's function technique\cite{Ihnatsenka} where the
electron- and spin interactions are included within the density functional
theory in the local spin density approximation. Our main findings can be
summarized as follows.

1) The magnetosubband structure and the density distribution in the
hard-wall quantum wire is qualitatively different from that one with a
smooth electrostatic confinement. In particularly, in the hard-wall wire a
deep triangular potential well of the width $\sim l_B$ is formed in the
vicinity of the wire boundary. The wave functions are strongly localized in
this well which leads to the increase of the electron density near the edges.

2) Because of the presence of the deep triangular well near the wire
boundaries, the subbands start to depopulate from the central region of the
wire and remain pinned in the well region until they are eventually pushed
up by an increasing magnetic field. This is in contrast to the case of a
smooth confinement where depopulation of the subbands starts from the edges
and extends towards the wire center as the magnetic field increases.

3) The spin polarization of electron density as a function of magnetic field
shows a pronounced double-loop pattern that can be related to the successive
depopulation of the magnetosubbands.

4) In contrast to the case of a smooth confinement, in the hard-wall wires
the compressible strips do not form in the vicinity of wire boundaries and a
spatial spin separation between spin-up and spin-down states near the edges
is absent.


\begin{acknowledgments}
S. I. acknowledges financial support from the Royal Swedish Academy of
Sciences and the Swedish Institute.
\end{acknowledgments}

\end{document}